\documentclass[12pt]{article}

\usepackage[T1]{fontenc}
\usepackage[latin1]{inputenc}
\usepackage{lmodern, csquotes}
\usepackage[english]{babel}

\usepackage[backend=biber,style=authoryear-comp,maxcitenames=1,sorting=nyt,sortcites=false]{biblatex}
\usepackage{amssymb, amsmath, dsfont, mathtools}
\usepackage{graphicx, diagbox, tikz}

\usepackage[normalem]{ulem}
\newcommand\soutpars[1]{\let\helpcmd\sout\parhelp#1\par\relax\relax}
\long\def\parhelp#1\par#2\relax{%
  \helpcmd{#1}\ifx\relax#2\else\par\parhelp#2\relax\fi%
}

\usepackage{pgfplots,pgfplotstable,booktabs,colortbl}
\usetikzlibrary{patterns}
\usepgfplotslibrary{fillbetween}
\pgfplotsset{compat=newest}
\pgfplotstableread{20241216_market_10_1000000.dat}\rhodat
\pgfplotstableread{20241216_market_10_1000000-vols.dat}\voldat

\definecolor{mylightyellow}{rgb}{1,1,.8}
\definecolor{mylightgreen}{rgb}{.8,1,.8}
\definecolor{mydarkorange}{RGB}{217,126,29}
\definecolor{mydarkred}{RGB}{178,34,34}
\definecolor{mydarkgreen}{RGB}{34,139,34}
\definecolor{mydarkblue}{RGB}{72,61,139}
\definecolor{mydarkyellow}{RGB}{218,165,32}

\usepackage{hyperref}
\hypersetup{
 colorlinks=true,breaklinks=true,
 urlcolor= mydarkblue,linkcolor= mydarkblue,citecolor= mydarkgreen,
 pdfauthor={Pallavicini, A.}
}

\DeclarePairedDelimiterX\QCond[2]{\{}{\}}{\,#1\,\delimsize\vert\,\mathopen{}#2\,}
\DeclarePairedDelimiterX\QUncond[1]{\{}{\}}{\,#1\,}
\DeclarePairedDelimiterX\ECond[2]{[}{]}{\,#1\,\delimsize\vert\,\mathopen{}#2\,}
\DeclarePairedDelimiterX\EUncond[1]{[}{]}{\,#1\,}

\newcommand{\Ex}[2]{\mathbb{E}_{#1}\EUncond*{#2}}
\newcommand{\ExT}[3]{\mathbb{E}_{#1}^{#2}\EUncond*{#3}}

\newcommand{\ExC}[3]{\mathbb{E}_{#1}\ECond*{#2}{#3}}
\newcommand{\ExCT}[4]{\mathbb{E}_{#1}^{#2}\ECond*{#3}{#4}}

\newcommand{\ind}[1]{1_{\{#1\}}}

\title{Pricing Quanto and Composite Contracts\\ with Local-Correlation Models}
\author{
Andrea Pallavicini\thanks{Intesa Sanpaolo, {\tt andrea.pallavicini@intesasanpaolo.com}}
}
\date{
\small First Version: November 29, 2023.  This version: \today
}

\begin{document}

\maketitle

\begin{abstract}

  Pricing composite and quanto contracts requires a joint model of both the underlying asset and the exchange rate. In this contribution, we explore the potential of local-correlation models to address the challenges of calibrating synthetic quanto forward contracts and composite options quoted in the market. Specifically, we design on-line calibration procedures for generic local and stochastic volatility models. The paper concludes with a numerical study assessing the calibration performance of these methodologies and comparing them to simpler approximations of the correlation structure.

\end{abstract}

\bigskip

\noindent {\bf AMS classification codes:} 65C05, 91G20, 91G60.

\newpage
{\small \tableofcontents}
\vfill
{\footnotesize \noindent The opinions here expressed are solely those of the author and do not represent in any way those of his employers.}
\newpage

\pagestyle{myheadings} \markboth{}{{\footnotesize Pallavicini, Pricing Quanto and Composite Contracts with LC Models}}

\section{Introduction}

Trading in foreign markets is often motivated by the desire to access a broader range of investment opportunities. However, strategies involving foreign assets are inherently subject to exchange-rate movements, which may not always align with investor objectives. Derivative contracts that maintain exposure to exchange-rate fluctuations are commonly referred to as composite contracts, whereas those designed to eliminate such exposure are known as quanto contracts.

An example of composite contracts is given by composite plain-vanilla options, which are designed for investors who seek to execute an option strategy on a foreign asset while fixing the strike price and receiving the payoff in their domestic currency. Conversely, an example of quanto contracts is given by quanto plain-vanilla options, which enable investors to mitigate exchange-rate fluctuations while trading in foreign markets. Also known as quantity-adjusting options, these cash-settled derivatives are denominated in a currency different from that of the underlying asset. This design allows investors to access foreign equities and other assets without assuming the risks associated with foreign currency movements. To achieve this, quanto plain-vanilla options establish a fixed exchange rate applied at contract settlement. The payoff is calculated by converting the value of the underlying asset into the payoff currency using this predetermined rate, allowing the holder to benefit from the performance of the underlying asset without exposure to exchange-rate movements.

Pricing composite and quanto contracts requires a joint model of both the underlying asset and the exchange rate. The model should calibrate liquid options on each risk factor along with available quotes on hybrid contracts. The literature on the subject usually focus on the evaluation of quanto options. To the best of our knowledge the first reference to quanto option pricing can be found in \textcite{Reiner1992}, where the Black-Scholes model is analyzed. Recent contributions focuses mainly on stochastic volatility models either with deterministic correlations, as in \textcite{Dimitroff2009,Giese2012,Fink2021}, or stochastic ones, as in \textcite{Branger2012,Teng2015}. A different line of research is followed by \textcite{Vong2014,Hong2020} which discuss local-volatility models with deterministic correlations. In particular, the latter one introduces Markov projection techniques to evaluate quanto options.

In this contribution we start from the analysis of \textcite{Mercurio2016,Guyon2017} on local-correlation models for multi-currency and basket options to address the problem of calibrating synthetic quanto forward contract quoted in the market. Local-correlation models rely on Markov projections to calibrate plain-vanilla quotes, which involve the calculation of forward expectations. Our main result is the design of a calibration procedure for synthetic quanto forward contracts which avoids such calculation, which usually requires a delicate fine tuning and greater computational costs. Then, we extend the model to include the calibration of composite options, and we hint at a possible joint calibration of quanto and composite contracts. As a last theoretical contribution, we consider stochastic volatility models for the underlying asset price and exchange rate, by showing that the calibration procedures naturally extends also to this setting. Finally, in the numerical section we compare two variants of our local-correlation model against a simpler approximation, which allows to guess the correlation structure directly from market data.

\medskip

The paper is organized as follows. In Section~\ref{sec:model} we introduce the local-correlation model, along with its extensions, and we describe how to calibrate the model to market data. In Section~\ref{sec:numerics} we discuss the numerical aspects of the calibration procedure for different choices of the correlation structure and we display the plain-vanilla volatilities implied by these choices. The paper concludes with two appendices that describe how market contracts are quoted and how to design a joint calibration.



\section{Model setup}
\label{sec:model}

We consider a probability space $\{\Omega,\mathcal{F},\mathbb{Q}\}$, where market information is described by the filtration $\mathcal{F}$ and the domestic risk-neutral pricing measure is $\mathbb{Q}$. The market quotes prices of a foreign asset $S^{\rm f}_t$ and an exchange rate $\chi_{{\rm f},t}$ converting prices expressed in foreign currency into priced in domestic currency\footnote{In this note superscripts indicate the currency, so that $S^{\rm f}_t$ means that the asset price is in currency $\rm f$, while subscripts indicate simply a dependency. Indeed, the exchange rate $\chi_{{\rm f},t}$ is a domestic asset.}. The numeraire of the domestic measure is the domestic bank account $B_t$. We assume that the market is quoting also a foreign bank account $B^{\rm f}_t$ allowing us to define the foreign risk-neutral measure $\mathbb{Q}^{\rm f}$ by means of the numeraire $\chi_{{\rm f},t} B^{\rm f}_t$. We always assume that interest rates are deterministic. 

\subsection{Local correlation}

First, we introduce a local-volatility dynamics for the asset price as given by
\begin{equation}
\label{eq:lvs}
 \frac{ds^{\rm f}_t}{s^{\rm f}_t} = \eta(t,s^{\rm f}_t) \,dW^{s,{\rm f}}_t
 ,\quad
 S^{\rm f}_t := s^{\rm f}_t F_0^{\rm f}(t) ,
\end{equation}
where $\eta$ is the local volatility function, and $W^{s,{\rm f}}_t$ is a standard Brownian motion under the foreign measure. $F_0^{\rm f}(t) := \ExT{0}{\rm f}{S^{\rm f}_t}$ is the forward price, and its value can be derived from market quotes. Details on the local volatility model can be found in \textcite{Dupire1994,Derman1994}.

Then, we introduce a local-volatility dynamics for the exchange rate as given by
\begin{equation}
\label{eq:lvx}
  \frac{dx_{{\rm f},t}}{x_{{\rm f},t}} = \psi(t,x_{{\rm f},t}) \,dW^x_t
 ,\quad
 \chi_{{\rm f},t} := x_{{\rm f},t} X_{{\rm f},0}(t) ,
\end{equation}
where $\psi$ is the local volatility function, and $W^x_t$ is a standard Brownian motion under the domestic measure. $X_{{\rm f},0}(t) := \Ex{0}{\chi_{{\rm f},t}}$ is the forward rate, and its value can be derived from market quotes.

The choice of the instantaneous correlation between the two Brownian motions is a kew element to describe quanto and composite contracts. Here, we start from the analysis of \textcite{Mercurio2016} on multi-currency options and \textcite{Guyon2017} on basket options, and we assume that the instantaneous correlation is a deterministic function of time and state. Thus, we write
\begin{equation}
  d \langle W^s,W^x \rangle_t = \rho(t,s^{\rm f}_t,x_{{\rm f},t}) \,dt.
\end{equation}

For later use we write the asset price dynamics under the domestic measure
\begin{equation}
 \frac{ds^{\rm f}_t}{s^{\rm f}_t} = - \rho(t,s^{\rm f}_t,x_{{\rm f},t})\eta(t,s^{\rm f}_t)\psi(t,x_{{\rm f},t}) \,dt + \eta(t,s^{\rm f}_t) \,dW^s_t .
\label{eq:quanto_price}
\end{equation}
Furthermore, we notice that the composite price $Z_t := \chi_{{\rm f},t} S^{\rm f}_t$ is a domestic asset with dynamics
\begin{equation}
 \frac{dz_t}{z_t} = \eta(t,s^{\rm f}_t) \,dW^s_t + \psi(t,x_{{\rm f},t}) \,dW^x_t
 ,\quad
 Z_t = z^{\rm f}_t X_{{\rm f},0}(t) F_0^{\rm f}(t) ,
\label{eq:composite_price}
\end{equation}
since we have that $\Ex{0}{Z_t} = X_{{\rm f},0}(t) F_0^{\rm f}(t)$.

We continue our analysis by discussing the calibration procedure for the local-correlation function. Then, at the end of this Section, we present the extension of the model to include stochastic volatilities. The calibration procedure can be extended also to this setting.

\subsection{Model calibration}

We assume that the market is quoting forward contracts and plain-vanilla options on the foreign asset and on the exchange rate. Thus, the forward price $F_0^{\rm f}(t)$ and forward exchange rate $X_{{\rm f},0}(t)$ can be derived from market quotes, as well as the local-volatility functions $\eta(t,s)$ and $\psi(t,x)$. We briefly discuss how to perform the calibration of local volatilities in the numerical section.

The only free parameter remains the local correlation $\rho(t,s,x)$, which we could calibrate, if we have at disposal market quotes of hybrid options sensitive to the correlation between the asset price and the exchange rate. The best candidates are quanto forward contracts and composite plain-vanilla options. We refer to the numerical section for a more detailed analysis on market contracts. In particular, quanto forward contracts are traded in the markets as a package of four plain-vanilla options. More details can be found in Section~\ref{sec:numerics}. Here, we stick on a simpler definition for ease of exposition.

Quanto forward contracts exchange at maturity a quantity of domestic currency against the value of the foreign asset converted in domestic currency with a unitary exchange rate. These contracts allow to calibrate the so-called ``quanto correlation'' defined as
\begin{equation}
  \gamma(t) := - \frac{1}{\sigma_S(t,1) \sigma_\chi(t,1) t} \,\log \frac{\Ex{0}{S^{\rm f}_t}}{F_0^{\rm f}(t)} ,
\end{equation}
where $\sigma(t,k)$ is the volatility of a plain-vanilla option of maturity $t$ and forward moneyness $k$. We notice that, if we model both the asset price and the exchange rate with a Black dynamics calibrated to at-the-money volatilities, than the correlation which allows to calibrate quanto forward contracts is, in fact, the quanto correlation. It is useful to define also the ``quanto correction'' as
\begin{equation}
\label{eq:qc}
  q(t) := e^{ - \gamma(t) \sigma_S(t,1) \sigma_\chi(t,1) t } = \Ex{0}{s_t} .
\end{equation}
The dynamics of the asset price, given by equation~\eqref{eq:quanto_price}, is calibrated to quanto correlations if the following constraint holds
\begin{equation}
 q(t) = 1 - \int_0^t \Ex{0}{s^{\rm f}_u \,\rho(u,s^{\rm f}_u,x_{{\rm f},u})\eta(u,s^{\rm f}_u)\psi(u,x_{{\rm f},u})} du ,
\end{equation}
or after taking the derivatives with respect to maturity
\begin{equation}
 \partial_t q(t) = - \Ex{0}{s^{\rm f}_t \,\rho(t,s^{\rm f}_t,x_{{\rm f},t})\eta(t,s^{\rm f}_t)\psi(t,x_{{\rm f},t})} .
\label{eq:quanto_correction}
\end{equation}

We remark that the above constraint ensures that quanto corrections are recovered by the model, but it does not determine a unique volatility surface for quanto or composite plain-vanilla options. Indeed, in Appendix~\ref{sec:joint} we hint at a possible strategy to perform a joint calibration to quanto corrections and composite options, if both available in the market.

We continue by analyzing composite plain-vanilla options. These contracts are plain-vanilla options on the composite price. As a first step, we introduce a local-volatility model for the composite price as given by
\begin{equation}
 \frac{d{\hat z}_t}{{\hat z}_t} = \phi(t,{\hat z}_t) \,dW^{\hat z}_t ,
\end{equation}
and calibrate the local-volatility function $\phi(t,{\hat z})$ to composite options. Then, the marginal distributions of ${\hat z}_t$ and $z_t$, the latter given by equation~\eqref{eq:composite_price}, are the same if the following constraint holds
\begin{equation}
 \phi^2(t,z) = \ExC{0}{\eta^2(t,s^{\rm f}_t) + \psi^2(t,x_{{\rm f},t}) + 2\rho(t,s^{\rm f}_t,x_{{\rm f},t})\eta(t,s^{\rm f}_t)\psi(t,x_{{\rm f},t})}{x_ts^{\rm f}_t=z} .
\label{eq:composite_lv}
\end{equation}
This results derives from the Gy{\"o}ngy Lemma presented in \textcite{Gyongy1986}.

We notice that we could employ the Gy{\"o}ngy Lemma also for the calibration of quanto corrections, as done in \textcite{Hong2020} with the specification of a local-drift model. However, in this way, we are forcing a stronger constraint than Equation~\eqref{eq:quanto_correction}, which is not needed for their calibration and leads to a more complex numerical implementation.

Usually either quanto correlations or composite options are quoted in the market. Thus, we devise two calibration procedures according to which quantity is at disposal. The other quantity can be implied by the model, once the local-correlation is calibrated. Nevertheless, we present in Appendix~\ref{sec:joint} also a strategy to achieve a joint calibration in case the quotes are at disposal.

\subsection{Calibration of quanto correlations}

In this sub-section we assume that the market is quoting synthetic forward quanto contracts, expressed in term of quanto correlations, and we wish to calibrate the model to these quotes. In Section~\ref{sec:numerics} we focus on this case to illustrate our numerical investigations, since this is the usual case occurring in the market.

If we assume that the local-correlation is only a function of time, we can write, by starting from equation~\eqref{eq:quanto_correction},
\begin{equation}
\label{eq:qc_calib_lv}
 \rho(t,s,x) \doteq \rho_{\rm LV}(t) := - \frac{\partial_t q(t)}{\Ex{0}{s^{\rm f}_t \,\eta(t,s^{\rm f}_t)\psi(t,x_{{\rm f},t})}} .
\end{equation}
Then, we can design a Monte Carlo simulation, where at each time step we calculate the correlation for the next time step by calculating the expectation as an average over the already simulated sample.

We can further approximate the previous equation, if we assume that the quanto corrections can be calculate by a Black-Scholes model, namely we can write
\begin{equation}
\label{eq:qc_calib_bs}
 \rho(t,s,x) \doteq \rho_{\rm BS}(t) := - \frac{\partial_t \log q(t)}{\sigma_S(t,1)\sigma_\chi(t,1)} \approx \rho_{\rm LV}(t) .
\end{equation}
In section~\ref{sec:numerics} we will investigate empirically on implied volatilities the impact of approximating local volatilities with at-the-money market volatilities.

Alternatively, without introducing any approximation, we can find the pathwise solution of equation~\eqref{eq:quanto_correction} and obtain the simpler, but potentially more irregular, local correlation function
\begin{equation}
\label{eq:qc_calib_lc}
  \rho(t,s,x) \doteq \rho_{\rm LC}(t,s,x) := - \frac{\partial_t \log q(t)}{\eta(t,s)\psi(t,x)} .
\end{equation}
where we use the relationship $q(t) = \Ex{0}{s_t}$ and the positivity of $s_t$.

For any choice of the local correlation we notice that the algorithm does not ensure that the resulting correlation is within the $[-1,1]$ range, so that we have to clip the result and check the quality of the calibration ex post.

\subsection{Calibration of composite options}

In this sub-section we assume that the market is quoting composite plain-vanilla options, and we wish to calibrate the model to these quotes. This is not the usual case occurring in the market, but we think that this analysis completes the discussion.

If we assume that the local-correlation is only a function of time and composite price, we can write, by starting from equation~\eqref{eq:composite_lv},
\begin{equation}
  \rho(t,s,x) \doteq \rho_{\rm LV2}(t,sx) := \frac{\phi^2(t,z) - \ExC{0}{\eta^2(t,s^{\rm f}_t) + \psi^2(t,x_{{\rm f},t})}{x_{{\rm f},t}s^{\rm f}_t=xs}}{2\,\ExC{0}{\eta(t,s^{\rm f}_t)\psi(t,x_{{\rm f},t})}{x_ts^{\rm f}_t=xs}} .
\end{equation}
Then, we can design a Monte Carlo simulation, where at each time step we calculate the correlation for the next time step by calculating the conditional expectations by means of a particle algorithm on the already simulated sample. This solution is similar to the one discussed in \textcite{Guyon2017} for basket options.

Alternatively, we can find a pathwise solution of equation~\eqref{eq:composite_lv} and obtain the simpler, but potentially more irregular, local-correlation function
\begin{equation}
  \rho(t,s,x) \doteq \rho_{\rm LC2}(t,s,x) = \frac{\phi^2(t,xs) - \eta^2(t,s) - \psi^2(t,x)}{2\,\eta(t,s)\psi(t,x)} ,
\end{equation}
which allows to perform a standard Monte Carlo simulation. This solution is similar to the one discussed in \textcite{Mercurio2016} for multi-currency options.

In both cases the algorithm does not ensure that the resulting correlation is within the $[-1,1]$ range, so that we have to clip the result and check the quality of the calibration ex post.
  
\subsection{Extension to stochastic volatilities}

We can formulate the problem with more general models for the underlying risk factors. Since the line of reasoning is the same of simpler model, we limit ourselves to derive the calibration equations corresponding to~\eqref{eq:quanto_correction} and~\eqref{eq:composite_lv}.

We could consider the following stochastic-local volatility (SLV) models\footnote{With a slight abuse of notation we use the same symbols for these SLV processes and the corresponding local-volatility processes defined in the previous sections.} for the asset price and the exchange rate under the domestic risk-neutral measure.
\begin{equation}
  \frac{ds^{\rm f}_t}{s^{\rm f}_t} = \ell_s(t,s^{\rm f}_t) \sqrt{v^s_t} \,dW^{s,{\rm f}}_t
  ,\quad
  \frac{dx_{{\rm f},t}}{x_{{\rm f},t}} = \ell_x(t,x_{{\rm f},t}) \sqrt{v^x_t} \,dW^x_t,
\end{equation}
where the deterministic leverage functions $\ell_s(t,s)$ and $\ell_x(t,x)$, and the processes $v^s_t$ and $v^t$ adapted to the market filtration, are regular enough to ensure a solution of the SDEs. 

Now, we assume that the instantaneous correlation between the two Brownian motions is a deterministic function of time and state, inclusive of the volatility processes, and it is given by
\begin{equation}
  d \langle W^s,W^x \rangle_t = \rho(t,s^{\rm f}_t,x_{{\rm f},t},v^s_t,v^x_t) \,dt.
\end{equation}

We can derive under these dynamics the calibration equations previously derived in the local-volatility case to implement the calibration of quanto and composite options. We start by considering the quanto correction. We can write
\begin{equation}
  \partial_t q(t) = - \Ex{0}{s^{\rm f}_t \,\rho(t,s^{\rm f}_t,x_{{\rm f},t},v^s_t,v^x_t) \ell_s(t,s^{\rm f}_t) \ell_x(t,x_{{\rm f},t}) \sqrt{v^s_tv^x_t}}.
\end{equation}
If we assume that the local-correlation is only a function of time, we can write
\begin{equation}
  \rho(t,s,x) \doteq \rho_{\rm LV3}(t) := - \frac{\partial_t q(t)}{\Ex{0}{s^{\rm f}_t \,\ell_s(t,s^{\rm f}_t) \ell_x(t,x_{{\rm f},t}) \sqrt{v^s_tv^x_t}}},
\end{equation}
or we can proceed by solving pathwise the constraints. On the other hand, if we consider the composite options, we have
\begin{multline}
  \phi^2(t,z) = \ExC{0}{\eta^2(t,s^{\rm f}_t) + \psi^2(t,x_{{\rm f},t})}{x_ts^{\rm f}_t=z_t} \\ + 2 \,\ExC{0}{\rho(t,s^{\rm f}_t,x_{{\rm f},t},v^s_t,v^x_t) \ell_s(t,s^{\rm f}_t) \ell_x(t,x_{{\rm f},t}) \sqrt{v^s_tv^x_t}}{x_ts^{\rm f}_t=z},
\end{multline}
where the first expectation is obtained by means of the Gy\"ongy lemma to ensure the calibration of plain-vanilla options, namely by imposing the relationships
\begin{equation}
  \eta^2(t,s) = \ell^2_s(t,s) \,\ExCT{0}{\rm f}{v^s_t}{s_t=s}
  ,\quad
  \psi^2(t,s) = \ell^2_x(t,s) \,\ExC{0}{v^x_t}{x_t=x}.
\end{equation}
If we assume that the local-correlation is only a function of time and composite price, we can write
\begin{equation}
  \rho(t,s,x) \doteq \rho_{\rm LV4}(t,sx) := \frac{\phi^2(t,xs) - \ExC{0}{\eta^2(t,s^{\rm f}_t) + \psi^2(t,x_{{\rm f},t})}{x_{{\rm f},t}s^{\rm f}_t=xs}}{2\,\ExC{0}{\ell_s(t,s^{\rm f}_t) \ell_x(t,x_{{\rm f},t}) \sqrt{v^s_tv^x_t}}{x_ts^{\rm f}_t=xs}},
\end{equation}
or we can proceed by solving pathwise the constraints.

\section{Numerical investigations}
\label{sec:numerics}

In this section we wish to calibrate the local-volatility version of the model in a real market case. We consider market quotes of plain-vanilla options on the DJ Euro Stoxx 50 index and synthetic quanto forward contracts on the same index paid in USD. We wish to calibrate these market data and to imply volatilities for quanto and composite plain-vanilla options in USD on the same index.

\subsection{Market data}

We observe call and put plain-vanilla options on DJ Euro Stoxx 50 index quoted in the Eurex market, while we use broker data to obtain information on EUR/USD plain-vanilla options. Forward prices of the index are derived by ensuring the put-call parities. The interest-rate curve for currency EUR is bootstrapped from ESTR-based interest-rate swaps, while the curve for currency USD is implied to match quoted exchange-rate forward swaps. All market data are observed on 16 December 2024. In this setting we recall that EUR is the foreign currency, while USD is the domestic currency.

Broker data are available for quanto correlations, while we are not able to find liquid quotes for composite options. Thus, we focus our investigations on calibrating quanto correlations and implying quanto and composite volatilities. We describe in Appendix~\ref{sec:broker} how broker data allows to reconstruct quanto correlations.

\subsection{Calibration of plain-vanilla options}

Plain-vanilla options on DJ Euro Stoxx 50 index and on EUR/USD are employed to calibrate the local-volatility functions introduced in equations~\eqref{eq:lvs} and~\eqref{eq:lvx}. We design a calibration procedure by following \textcite{Nastasi2020} with the proper modifications to adapt the algorithm to equity indices and exchange rates. We refer to such paper for numerical details.

We briefly describe the procedure for an equity index, whose dynamics is given by equations~\eqref{eq:lvs}. A similar argument holds for exchange rates. We consider that European-style plain-vanilla options are quoted on the market in term of Black-Scholes volatilities which we identify as
\begin{equation}
  \{\sigma^{\rm mkt}(t_i,k_i^j) : i\in[1,N], j\in[1,M_i]\},\quad
  \sigma^{\rm mkt}_{ij} := \sigma^{\rm mkt}(t_i,k_i^j).
\end{equation}
We notice that strike prices may depend on maturities. For instance, the foreign-exchange market quotes options in term of Black-Scholes Delta leading to a different set of strikes for each maturity. Model calibration can be achieved by specifying a particular non-parametric form for the local-volatility function
\begin{equation}
  \eta(t,k) := \zeta(t,k;\{\eta(t_i,k_i^j)\}),\quad
  \eta_{ij} := \eta(t_i,k_i^j),
\end{equation}
where $\zeta$ is a function used to interpolate and extrapolate the local-volatility function. We select a piece-wise-constant function in time, while in strike we choose a cubic monotone spline for interpolation and constant function for extrapolation. Then, the calibration procedure is given by the following fixed-point scheme
\begin{enumerate}
  \item We set the nodes of the local-volatility function $\eta^{(0)}$ to be equal to the market volatilities.
  \item We calculate the spot volatilities $\sigma^{(0)}$ implied from the model when the local-volatility is interpolated on nodes $\eta^{(0)}$.
  \item We compare $\sigma^{(0)}$ with $\sigma^{\rm mkt}$, and, if the distance is within a given threshold we stop the algorithm. Otherwise the algorithm proceeds to the next step.
  \item We adjust the value of the local-volatility nodes by taking into account the first-order corrections for small time-to-maturities and near-ATM strikes to volatility level and skew derived in \textcite[Proposition 4.2]{Nastasi2020}, as given by
  \begin{equation}
    \eta^{(1)}_{ij} = \eta^{(0)}_{ij} \frac{\sigma^{\rm mkt}_{ij_{\rm atm}}}{\sigma^{(0)}_{ij_{\rm atm}}} + 2 \left( \frac{\partial \sigma^{\rm mkt}_{ij}}{\partial k} - \frac{\partial \sigma^{(0)}_{ij}}{\partial k} \right) \Delta k_j \ind{j\neq j_{\rm atm}}
  \end{equation}
  where $j_{\rm atm}$ is the strike index referring to at-the-money options.
  \item We repeat the algorithm from the third step with $\eta^{(1)}$. Then, the iteration is repeated until convergence, or until a maximum number of iterations is reached.
\end{enumerate}

With few iterations of the algorithm is possible to achieve a precision in term of absolute error between implied and market volatilities well below the basis point.

\subsection{Calibration of quanto correlations}

We selected a local-volatility dynamics both for the equity index and the exchange rate. However, we can choose among three different strategies to calibrate quanto correlations:
\begin{enumerate}
  \item a simple model based on a Black-Scholes approximation, given by equation~\eqref{eq:qc_calib_bs}, which we identify as ``BS'';
  \item a time-dependent correlation model, given by equation~\eqref{eq:qc_calib_lv}, which we identify as ``LV'';
  \item a local-correlation model, given by equation~\eqref{eq:qc_calib_lc}, which we identify as ``LC''. 
\end{enumerate} 
All these strategies are on-line calibrations, meaning that we compute the relevant parameters within the same simulation we use when pricing.

In all strategies the relevant issue is the knowledge of quanto corrections at all simulation times. In the market we can retrieve information only for a limited number of maturity dates, so that we need to mark a term structure for quanto corrections.

In our numerical investigations we introduce a non-parametric form for quanto correlations (we do not start directly from quanto corrections, because we think the former ones have a more clear financial meaning). We write
\begin{equation}
  \gamma(t) := \xi(t;\{\gamma(t_i)\}), \quad \gamma_i := \gamma(t_i),
\end{equation}
where $\xi$ is a function used to interpolate and extrapolate the quanto correlation function. We select a cubic monotone spline for interpolation, and constant function for extrapolation.

Node values of the quanto correlation function can be read directly from broker quotes, as described in appendix~\ref{sec:broker}. Then, quanto corrections can be calculate by inverting equation~\eqref{eq:qc}. We notice that this procedure depends on the at-the-money volatilities $\sigma_S(t,1)$ and $\sigma_\chi(t,1)$, which can be calculated for any time $t$ by solving the Dupire equation once the local volatility models of each rick factor are calibrated.

\begin{table}
\begin{center}
  \pgfplotstableset{
    create on use/quo_mid/.style={create col/expr={(\thisrow{quo_ask}+\thisrow{quo_bid})*5000}}
  }
  \scalebox{0.95}{
  \pgfplotstabletypeset[
    fixed, fixed zerofill, precision=1, set thousands separator={},
    every head row/.style={output empty row},
    every head row/.append style={%
      after row  = \midrule,
      before row = \toprule
                    \multicolumn{10}{l}{\bf\boldmath Quanto forward contracts -- $\Gamma(T)$} \smallskip\\
                    \midrule
                    {\bf Maturity} & {\bf Bid} & {\bf Mid} & {\bf Ask}
                    & \multicolumn{2}{c|}{{\bf BS}}
                    & \multicolumn{2}{c|}{{\bf LV}}
                    & \multicolumn{2}{c}{{\bf LC}}
                    \\},
    every last row/.style={after row=\bottomrule},
    columns={label,quo_ask,quo_mid,quo_bid,quo_bs,quo_bs_e,quo_lv,quo_lv_e,quo_lc,quo_lc_e},
    columns/label/.style={string type,postproc cell content/.style={@cell content={\textsc{##1}}},column type/.add={}{|}},
    columns/quo_bid/.style={preproc/expr={10000*##1}},
    columns/quo_ask/.style={preproc/expr={10000*##1}},
    columns/quo_bs/.style={column type=r, column type/.add={|}{@{$\,\pm\,$}}, preproc/expr={10000*##1}},
    columns/quo_bs_e/.style={column type=l,preproc/expr={20000*##1}},
    columns/quo_lv/.style={column type=r, column type/.add={|}{@{$\,\pm\,$}},preproc/expr={10000*##1}},
    columns/quo_lv_e/.style={column type=l,preproc/expr={20000*##1}},
    columns/quo_lc/.style={column type=r, column type/.add={|}{@{$\,\pm\,$}},preproc/expr={10000*##1}},
    columns/quo_lc_e/.style={column type=l,preproc/expr={20000*##1}},
  ]{\rhodat}}
\end{center}
\caption{Prices in basis points of quanto forward contracts quoted by the market on 16 December 2024. The first column is the maturity of the contract. The next three columns the market prices, the last three columns the model implied prices. Statistical uncertainties are at $95\%$ of confidence level.} 
\label{tab:quote}
\end{table}

\begin{table}
\begin{center}
  \pgfplotstableset{
    create on use/rho_mid/.style={create col/expr={(\thisrow{rho_ask}+\thisrow{rho_bid})*50}}
  }
  \scalebox{0.95}{
  \pgfplotstabletypeset[
    fixed, fixed zerofill, precision=1, set thousands separator={},
    every head row/.style={output empty row},
    every head row/.append style={%
      after row  = \midrule,
      before row = \toprule
                    \multicolumn{10}{l}{\bf\boldmath Quanto correlations -- $\gamma(T)$} \smallskip\\
                    \midrule
                    {\bf Maturity} & {\bf Bid} & {\bf Mid} & {\bf Ask}
                    & \multicolumn{2}{c|}{{\bf BS}}
                    & \multicolumn{2}{c|}{{\bf LV}}
                    & \multicolumn{2}{c}{{\bf LC}}
                    \\},
    every last row/.style={after row=\bottomrule},
    columns={label,rho_ask,rho_mid,rho_bid,rho_bs,rho_bs_e,rho_lv,rho_lv_e,rho_lc,rho_lc_e},
    columns/label/.style={string type,postproc cell content/.style={@cell content={\textsc{##1}}},column type/.add={}{|}},
    columns/rho_bid/.style={preproc/expr={100*##1}},
    columns/rho_ask/.style={preproc/expr={100*##1}},
    columns/rho_bs/.style={column type=r, column type/.add={|}{@{$\,\pm\,$}}, preproc/expr={100*##1}},
    columns/rho_bs_e/.style={column type=l,preproc/expr={200*##1}},
    columns/rho_lv/.style={column type=r, column type/.add={|}{@{$\,\pm\,$}}, preproc/expr={100*##1}},
    columns/rho_lv_e/.style={column type=l,preproc/expr={200*##1}},
    columns/rho_lc/.style={column type=r, column type/.add={|}{@{$\,\pm\,$}}, preproc/expr={100*##1}},
    columns/rho_lc_e/.style={column type=l,preproc/expr={200*##1}},
  ]{\rhodat}}
\end{center}
\caption{Quanto correlations in percentage implied by market data on 16 December 2024. The first column is the maturity of the corresponding quanto forward contract. The next three columns the market implied correlations, the last three columns the model implied correlations. Statistical uncertainties are at $95\%$ of confidence level.}
\label{tab:rho}
\end{table}

We show in Table~\ref{tab:quote} the prices of quanto forward contracts quoted by the market on 16 December 2024, along with the ones implied by the three strategies just discussed when calibrating to mid market data. The definition of the contract is given in Appendix~\ref{sec:broker}. The corresponding implied quanto correlations are shown in Table~\ref{tab:rho}. All calculations are performed with $10^6$ Monte Carlo simulations and statistical uncertainties are at $95\%$ of confidence level. We can see that the LV and LC strategies lead to consistent and very similar behaviors for all maturity dates, while the BS approximation, as expected, departs from them. Yet, it is worth noticing that the approximation is still compatible with the market bid-ask spreads.

\subsection{Implied volatility of quanto and composite options}

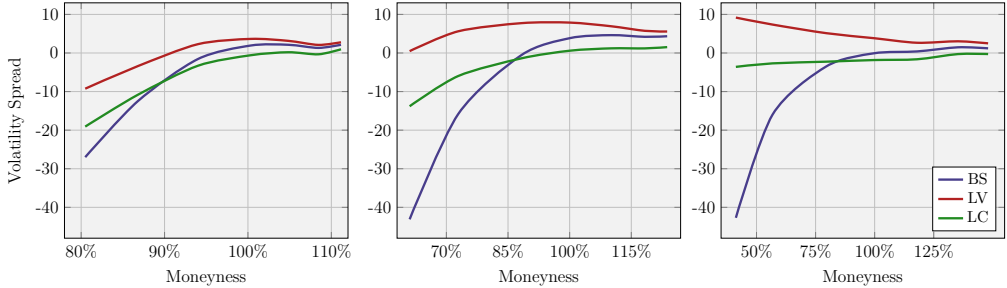
\begin{figure}
\scalebox{0.55}{
\begin{tikzpicture}
  \begin{axis}[set layers, mark layer=axis background,
              xlabel=Moneyness,
              ylabel=Volatility Spread,
              xmin=0.78, xmax=1.12,
              xtick={0.80,0.90,1.00,1.10},
              xticklabels={80\%,90\%,100\%,110\%},
              ymin=-48, ymax=13,
              ytick={-40,-30,-20,-10,0,10},
              yticklabels={-40,-30,-20,-10,0,10},
              grid=major,
              legend style={legend pos=south east},
              axis background/.style={fill=gray!10}]
  \addplot [color=mydarkblue,ultra thick,smooth] table [y expr={10000*(\thisrow{bq3}-\thisrow{v3})},x=m3] from \voldat;
  \addplot [color=mydarkred,ultra thick,smooth] table [y expr={10000*(\thisrow{vq3}-\thisrow{v3})},x=m3] from \voldat;
  \addplot [color=mydarkgreen,ultra thick,smooth] table [y expr={10000*(\thisrow{cq3}-\thisrow{v3})},x=m3] from \voldat;
  \end{axis}
\end{tikzpicture}
\begin{tikzpicture}
  \begin{axis}[set layers, mark layer=axis background,
              xlabel=Moneyness,
              xmin=0.58, xmax=1.27,
              xtick={0.70,0.85,1.00,1.15},
              xticklabels={70\%,85\%,100\%,115\%},
              ymin=-48, ymax=13,
              ytick={-40,-30,-20,-10,0,10},
              yticklabels={-40,-30,-20,-10,0,10},
              grid=major,
              legend style={legend pos=south east},
              axis background/.style={fill=gray!10}]
  \addplot [color=mydarkblue,ultra thick,smooth] table [y expr={10000*(\thisrow{bq5}-\thisrow{v5})},x=m5] from \voldat;
  \addplot [color=mydarkred,ultra thick,smooth] table [y expr={10000*(\thisrow{vq5}-\thisrow{v5})},x=m5] from \voldat;
  \addplot [color=mydarkgreen,ultra thick,smooth] table [y expr={10000*(\thisrow{cq5}-\thisrow{v5})},x=m5] from \voldat;
  \end{axis}
\end{tikzpicture}
\begin{tikzpicture}
  \begin{axis}[set layers, mark layer=axis background,
              xlabel=Moneyness,
              xmin=0.35, xmax=1.55,
              xtick={0.50,0.75,1.00,1.25},
              xticklabels={50\%,75\%,100\%,125\%},
              ymin=-48, ymax=13,
              ytick={-40,-30,-20,-10,0,10},
              yticklabels={-40,-30,-20,-10,0,10},
              grid=major,
              legend style={legend pos=south east},
              axis background/.style={fill=gray!10}]
  \addplot [color=mydarkblue,ultra thick,smooth] table [y expr={10000*(\thisrow{bq7}-\thisrow{v7})},x=m7] from \voldat;
  \addplot [color=mydarkred,ultra thick,smooth] table [y expr={10000*(\thisrow{vq7}-\thisrow{v7})},x=m7] from \voldat;
  \addplot [color=mydarkgreen,ultra thick,smooth] table [y expr={10000*(\thisrow{cq7}-\thisrow{v7})},x=m7] from \voldat;
  \legend{{\small BS},{\small LV},{\small LC}}
  \end{axis}
\end{tikzpicture}}
\caption{Spread in basis points between implied volatility of quanto options and market volatility of plain-vanilla options on 16 December 2024 for three different maturity dates $T$: \textsc{Mar25}, \textsc{Dec26} and \textsc{Dec27}. On the $x$-axis the forward moneyness, namely $K/F_0^{\rm f}(T)$.}
\label{fig:volq}
\end{figure}

We conclude the numerical investigation by calculating the implied volatilities for quanto and composite options with the three strategies discussed in the previous Sub-sections. In Figure~\ref{fig:volq} we show the difference between the implied volatility of quanto plain-vanilla options in USD on DJ Euro Stoxx 50 and the market volatility of plain-vanilla options in EUR on the same index, both calculated on maturities and forward moneyness ($K/F_0^{\rm f}(T)$) quoted by the market. In the pictures we use an interpolating smooth line to present the results. We can see that the three strategies produce different quanto volatilities: a difference of up to ten basis points between the strategies LV and LC on all maturities and strikes, while a greater difference is marked by the strategy BS. We explain the small values of the spreads, for all the strategies, by noticing that for flat volatility smiles the spread is null, so that we cannot expect large deviations when we consider market smiles. On the other hand, the BS case shows a greater difference because of its approximation in calculating quanto corrections which has the effect of changing the moneyness of the option, leading to a greater effect for small strikes since equity skews are greater for smaller strikes.

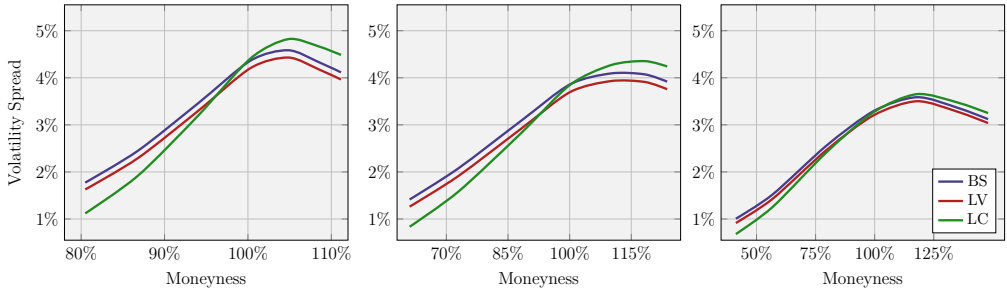
\begin{figure}
\scalebox{0.55}{
\begin{tikzpicture}
  \begin{axis}[set layers, mark layer=axis background,
              xlabel=Moneyness,
              ylabel=Volatility Spread,
              xmin=0.78, xmax=1.12,
              xtick={0.80,0.90,1.00,1.10},
              xticklabels={80\%,90\%,100\%,110\%},
              ymin=55, ymax=555,
              ytick={100,200,300,400,500},
              yticklabels={1\%,2\%,3\%,4\%,5\%},
              grid=major,
              legend style={legend pos=south east},
              axis background/.style={fill=gray!10}]
  \addplot [color=mydarkblue,ultra thick,smooth] table [y expr={10000*(\thisrow{bc3}-\thisrow{v3})},x=m3] from \voldat;
  \addplot [color=mydarkred,ultra thick,smooth] table [y expr={10000*(\thisrow{vc3}-\thisrow{v3})},x=m3] from \voldat;
  \addplot [color=mydarkgreen,ultra thick,smooth] table [y expr={10000*(\thisrow{cc3}-\thisrow{v3})},x=m3] from \voldat;
  \end{axis}
\end{tikzpicture}
\begin{tikzpicture}
  \begin{axis}[set layers, mark layer=axis background,
              xlabel=Moneyness,
              xmin=0.58, xmax=1.27,
              xtick={0.70,0.85,1.00,1.15},
              xticklabels={70\%,85\%,100\%,115\%},
              ymin=55, ymax=555,
              ytick={100,200,300,400,500},
              yticklabels={1\%,2\%,3\%,4\%,5\%},
              grid=major,
              legend style={legend pos=south east},
              axis background/.style={fill=gray!10}]
  \addplot [color=mydarkblue,ultra thick,smooth] table [y expr={10000*(\thisrow{bc5}-\thisrow{v5})},x=m5] from \voldat;
  \addplot [color=mydarkred,ultra thick,smooth] table [y expr={10000*(\thisrow{vc5}-\thisrow{v5})},x=m5] from \voldat;
  \addplot [color=mydarkgreen,ultra thick,smooth] table [y expr={10000*(\thisrow{cc5}-\thisrow{v5})},x=m5] from \voldat;
  \end{axis}
\end{tikzpicture}
\begin{tikzpicture}
  \begin{axis}[set layers, mark layer=axis background,
              xlabel=Moneyness,
              xmin=0.35, xmax=1.55,
              xtick={0.50,0.75,1.00,1.25},
              xticklabels={50\%,75\%,100\%,125\%},
              ymin=55, ymax=555,
              ytick={100,200,300,400,500},
              yticklabels={1\%,2\%,3\%,4\%,5\%},
              grid=major,
              legend style={legend pos=south east},
              axis background/.style={fill=gray!10}]
  \addplot [color=mydarkblue,ultra thick,smooth] table [y expr={10000*(\thisrow{bc7}-\thisrow{v7})},x=m7] from \voldat;
  \addplot [color=mydarkred,ultra thick,smooth] table [y expr={10000*(\thisrow{vc7}-\thisrow{v7})},x=m7] from \voldat;
  \addplot [color=mydarkgreen,ultra thick,smooth] table [y expr={10000*(\thisrow{cc7}-\thisrow{v7})},x=m7] from \voldat;
  \legend{{\small BS},{\small LV},{\small LC}}
  \end{axis}
\end{tikzpicture}
}
\caption{Spread in percentage between implied volatility of composite options and market volatility of plain-vanilla options on 16 December 2024 for three different maturity dates $T$: \textsc{Mar25}, \textsc{Dec26} and \textsc{Dec27}. On the $x$-axis the forward moneyness, namely $K/F_0^{\rm f}(T)$.}
\label{fig:volc}
\end{figure}

We continue the analysis with Figure~\ref{fig:volc}, where we show the difference between the implied volatility of composite plain-vanilla options in USD on DJ Euro Stoxx 50 and the market volatility of plain-vanilla options in EUR on the same index, both calculated on maturities and forward moneyness ($K/F_0^{\rm f}(T)$) quoted by the market. In the pictures we use an interpolating smooth line to present the results. Again we can see that the three strategies produce different quanto volatilities, but this time the BS and LV strategies lead to more similar results than the LC one, particularly for short maturities. In this case the approximation of the BS strategy does not alter the moneyness of the options, since the composite forward price does not depend on correlations, leading to values closer to the LV strategy. The difference between the LV and LC strategies is a measure of model risk. Indeed, since the market specify only partially the joint distribution of asset price and exchange rate, we can perfectly calibrate different models to equity and exchange rate markets, along with quanto corrections, and still obtain different values for hybrid products.

\section{Conclusion and further developments}

We presented a local-correlation model for foreign asset dynamics along with a calibration procedure to synthetic quanto forward contracts and composite options. We described also the model extension to a stochastic volatility setting. In the numerical part we investigated the calibration performance to real market data and the volatility smiles implied by the model under different choices of the correlation structure. In particular, we obtained an on-line calibration procedure to synthetic quanto forward contracts which does not involve the calculation of forward expectations. Moreover, we presented a simpler deterministic approximation of the correlation structure able to fit market data. A possible development of the present work is studying a calibration procedure for quanto plain-vanilla options and extending the model to a multiple assets.

\printbibliography

\appendix

\section{Synthetic forward contracts}
\label{sec:broker}

We describe in more detail how quanto corrections are available in the market, since it is a less standard product than plain-vanilla options or interest-rate swaps, the other contract used in this notes. Broker data usually contains information in term of the difference in price between the synthetic quanto forward contract and the synthetic forward contract. As a first step let us define at-the-money plain-vanilla call and put options prices as given by
\begin{equation}
  C(T) := \ExT{0}{\rm f}{ \left( \frac{S^{\rm f}_T}{S^{\rm f}_0} - 1 \right)^{\!\!+} \frac{B^{\rm f}_0}{B^{\rm f}_T} },\;
  P(T) := \ExT{0}{\rm f}{ \left( 1 - \frac{S^{\rm f}_T}{S^{\rm f}_0} \right)^{\!\!+} \frac{B^{\rm f}_0}{B^{\rm f}_T} },
\end{equation}
where we recall that $B^{\rm f}_t$ is the bank account in foreign currency and the expectations are taken under the risk-neutral measure in such currency. The, we define the quantized version of the previous contracts, given by
\begin{equation}
  C^q(T) := \Ex{0}{ \left( \frac{S^{\rm f}_T}{S^{\rm f}_0} - 1 \right)^{\!\!+} \frac{B_0}{B_T} },\;
  P^q(T) := \Ex{0}{ \left( 1 - \frac{S^{\rm f}_T}{S^{\rm f}_0} \right)^{\!\!+} \frac{B_0}{B_T} },
\end{equation}
where we recall that $B_t$ is the bank account in domestic currency and the expectations are taken under the risk-neutral measure in such currency. Finally we define broker quotes as
\begin{equation}
  \Gamma(T) := C^q(T) - P^q(T) - C(T) + P(T),
\end{equation}
which can be related to quanto correlations by using equation~\eqref{eq:qc}, leading to
\begin{equation}
  \Gamma(T) = \left(\frac{F^{\rm f}_0(T)}{S^{\rm f}_0} \,e^{\gamma(T) \sigma_S(T,1) \sigma_\chi(T,1) T} - 1\right)  \frac{B_0}{B_T} - \left(\frac{F^{\rm f}_0(T)}{S^{\rm f}_0} - 1\right)  \frac{B^{\rm f}_0}{B^{\rm f}_T}.
\end{equation}
The above equation can be used to derive quanto correlations $\gamma(T)$ from broker quotes $\Gamma(T)$ for each maturity date $T$.

\section{Joint calibration}
\label{sec:joint}

Within our modelling framework it is possible to design, in line of principle, a joint calibration algorithm for quanto corrections and composite plain-vanilla options, in case both prices are available in the market.

In former works in the literature only quanto corrections are usually considered, so that this problem is not discussed. In stochastic correlation models composite options could be added to the calibration procedure if the models are enough flexible to accommodate the quotes.

Here, we can achieve a joint calibration thanks to the flexibility of the local-correlation function. As a first step we introduce the function
\begin{equation}
\label{eq:theta}
  \theta(t,s,x) := \rho(t,s,x) \eta(t,s) \psi(t,x),
\end{equation}
so that we can write the relevant calibration equations~\eqref{eq:quanto_correction} and~\eqref{eq:composite_lv} in term of $\theta$ as given by
\begin{equation}
  \Ex{0}{s^{\rm f}_t \left( \partial_t \log q(t) + \theta(t,s^{\rm f}_t,x_{{\rm f},t}) \right)} = 0
\end{equation}
and
\begin{equation}
  \ExC{0}{\phi^2(t,s^{\rm f}_t x_{{\rm f},t}) - \eta^2(t,s^{\rm f}_t) - \psi^2(t,x_{{\rm f},t}) - 2\,\theta(t,s^{\rm f}_t,x_{{\rm f},t})}{x_{{\rm f},t} s^{\rm f}_t=z} = 0.
\end{equation}
Then, we split the function $\theta$ into two terms
\begin{equation}
  \theta(t,s,x) = \theta_z(t,sx) + \theta_r(t,s,x),
\end{equation}
where $\theta_z$ is defined as
\begin{equation}
  \theta_z(t,z) := \ExC{0}{\theta(t,s^{\rm f}_t,x_{{\rm f},t})}{x_{{\rm f},t} s^{\rm f}_t=z}
\end{equation}
and $\theta_r$ is a reminder defined once $\theta$ and $\theta_z$ are given. We can substitute the expression for $\theta$ in the second calibration equation to obtain an expression for $\theta_z$ which depends only on quantity that can be calibrated on the market. Indeed, we get
\begin{equation}
  \theta_z(t,z) = \frac{1}{2} \,\ExC{0}{\phi^2(t,s^{\rm f}_t x_{{\rm f},t}) - \eta^2(t,s^{\rm f}_t) - \psi^2(t,x_{{\rm f},t})}{x_{{\rm f},t} s^{\rm f}_t=z}.
\end{equation}
On the other hand, the reminder $\theta_r$ is a function that must satisfy only the constraint given by the first calibration equation, namely
\begin{equation}
\label{eq:theta_calibration}
  \Ex{0}{s^{\rm f}_t \,\theta_r(t,s^{\rm f}_t,x_{{\rm f},t})} = - \partial_t q(t) - \Ex{0}{s^{\rm f}_t \,\theta_z(t,x_{{\rm f},t} s^{\rm f}_t)},
\end{equation}
and the structural constraint coming from the definition of the function itself, which is given by
\begin{equation}
\label{eq:theta_structural}
  \ExC{0}{\theta_r(t,s^{\rm f}_t,x_{{\rm f},t})}{x_{{\rm f},t} s^{\rm f}_t=z} = 0.
\end{equation}

Hence, we get that the local correlation can be calculated, via Equation~\eqref{eq:theta}, from a function $\theta_r$ representing a non-trivial dependency between asset price and exchange rate, which is not completely determined by market data. There are infinite functions with the properties required for $\theta_r$. For instance, we can define $\theta_r$ as
\begin{equation}
  \theta_r(t,s,x) := \frac{\kappa(t) (1-s)e^{-s}}{p_{s^{\rm f}_t,x_{{\rm f},t}}(s,x)},
\end{equation}
where $p_{s^{\rm f}_t,x_{{\rm f},t}}$ is the joint probability density of asset price and exchange rate at time $t$, and $\kappa$ is a deterministic function of time. We can check that Equation~\eqref{eq:theta_structural} is satisfied by an explicit calculation, while from Equation~\eqref{eq:theta_calibration} we can calibrate $\kappa$. Indeed, we get
\begin{equation}
  \kappa(t) = \partial_t q(t) + \Ex{0}{s^{\rm f}_t \,\theta_z(t,x_{{\rm f},t} s^{\rm f}_t)}.
\end{equation}
However, in practice, a proper definition of $\theta_r$ is not trivial to find, and we leave it for a future work, since we must ensure that the local correlation, obtained via Equation~\eqref{eq:theta}, remains bounded in the $[-1,1]$ range.

\end{document}